\documentclass[aps,prx,twocolumn,eqsecnum,nofootinbib,superscriptaddress]{revtex4-2}
\pdfoutput=1 

\usepackage{float}
\usepackage{amsmath}
\usepackage{amssymb}
\usepackage{graphicx}
\usepackage{mathrsfs}
\usepackage{color}
\usepackage{hyperref}
\usepackage{lipsum}
\usepackage[procnames]{listings}
\usepackage{layouts}
\usepackage{epstopdf}
\usepackage{dsfont}
\usepackage{bbm}
\usepackage{ulem}
\usepackage{bbold}
\usepackage{mathtools}
\newcommand{\bra}[1]{{\left\langle{#1}\right\vert}}
\newcommand{\ket}[1]{{\left\vert{#1}\right\rangle}}
\newcommand{\expec}[1]{{\langle{#1}\rangle}}

\DeclareUnicodeCharacter{2009}{\,} 
\newcommand{\nq}{\hbar}
\begin{document}

\title{A Hybrid Approach to Mitigate Errors in Linear Photonic\\ Bell-State Measurement for Quantum Interconnects}

\author{Beate E. Asenbeck} 
\affiliation{Laboratoire Kastler Brossel, Sorbonne Universit\'e, CNRS, ENS-Universit\'e PSL, Coll\`ege de France, 4 Place Jussieu, 75005 Paris, France}
\author{Akito Kawasaki}
\affiliation{Department of Applied Physics, School of Engineering, The University of Tokyo, 7-3-1 Hongo, Bunkyo-ku, Tokyo 113-8656, Japan}
\author{Ambroise Boyer}
\affiliation{Laboratoire Kastler Brossel, Sorbonne Universit\'e, CNRS, ENS-Universit\'e PSL, Coll\`ege de France, 4 Place Jussieu, 75005 Paris, France}
\author{Tom Darras }
\affiliation{Welinq, 14 rue Jean Mac\'e, 75011 Paris, France}
\author{Alban Urvoy}
\affiliation{Laboratoire Kastler Brossel, Sorbonne Universit\'e, CNRS, ENS-Universit\'e PSL, Coll\`ege de France, 4 Place Jussieu, 75005 Paris, France}
\author{Akira Furusawa}
\affiliation{Department of Applied Physics, School of Engineering, The University of Tokyo, 7-3-1 Hongo, Bunkyo-ku, Tokyo 113-8656, Japan}
\affiliation{Optical Quantum Computing Research Team, RIKEN Center for Quantum Computing, 2-1 Hirosawa, Wako, Saitama 351-0198, Japan}
\author{Julien Laurat}
\email[email: ]{julien.laurat@sorbonne-universite.fr}
\affiliation{Laboratoire Kastler Brossel, Sorbonne Universit\'e, CNRS, ENS-Universit\'e PSL, Coll\`ege de France, 4 Place Jussieu, 75005 Paris, France}

\begin{abstract}
Optical quantum information processing critically relies on Bell-state measurement, a ubiquitous operation for quantum communication and computing. Its practical realization involves the interference of optical modes and the detection of a single photon in an indistinguishable manner. Yet, in the absence of efficient photon-number resolution capabilities, errors arise from multi-photon components, decreasing the overall process fidelity. Here, we introduce a hybrid detection scheme for Bell-state measurement, leveraging both on-off single-photon detection and quadrature conditioning via homodyne detection. We derive explicit fidelities for quantum teleportation and entanglement swapping processes employing this strategy, demonstrating its efficacy. We also compare with photon-number resolving detectors and find a strong advantage of the hybrid scheme in a wide range of parameters. This work provides a new tool for linear optics schemes, with applications to quantum state engineering and quantum interconnects.
\end{abstract}

\date{\today}

\maketitle

\section{Introduction}

Quantum networks and interconnects \cite{Awschalom2021} rely heavily on the Bell-state measurement (BSM), a cornerstone operation in linear-optical quantum information processing. The Bell-state measurement is a key step for quantum teleportation \cite{Bennet1993,Braunstein1995,Lutkenhaus1999,Pirandola2015} and more generally at the core of quantum connectivity \cite{Kimble2008,Wehner2018} by enabling entanglement heralding between distant systems \cite{Cabrillo1999,DeRiedmatten,Laurat,Slodicka2013,Hanson2018} and long-distance quantum communication via entanglement swapping \cite{Pan1998, Laurat2007} performed at each node of a network \cite{Briegel1998, DLCZ2001,Azuma2023}. BSM also serves as a central operation in quantum computing, either for modular approaches where small quantum processors can be interconnected \cite{Awschalom2021} or for computing architectures such as measurement- or fusion-based approaches \cite{Raussendorf2013,Bartolucci2023,Hilaire2023}.

Bell-state measurements rely on the interference of two optical modes followed by photon detection \cite{Braunstein1995}. Central to this operation is the capability to detect a single-photon state. However, a two-photon component is present in most photon creation processes, and inherently in the BSM as more than one photon can interfere on the beamsplitter. Commonly used on-off detectors, such as avalanche photodiodes or superconducting detectors, cannot distinguish photon numbers, and are thus unable to discriminate one- from two-photon contributions \cite{Hadfield2009}. This two-photon error is a strong limitation for achieving high-fidelity Bell-state measurements. A potential alternative to on-off detectors are superconducting photon-number resolving detectors \cite{Lita2022}, which are still under intense development and of limited availability, or a large number of multiplexed on-off detectors that would require extremely high efficiencies. 

\begin{figure}[b!]
\includegraphics[width=0.95\columnwidth]{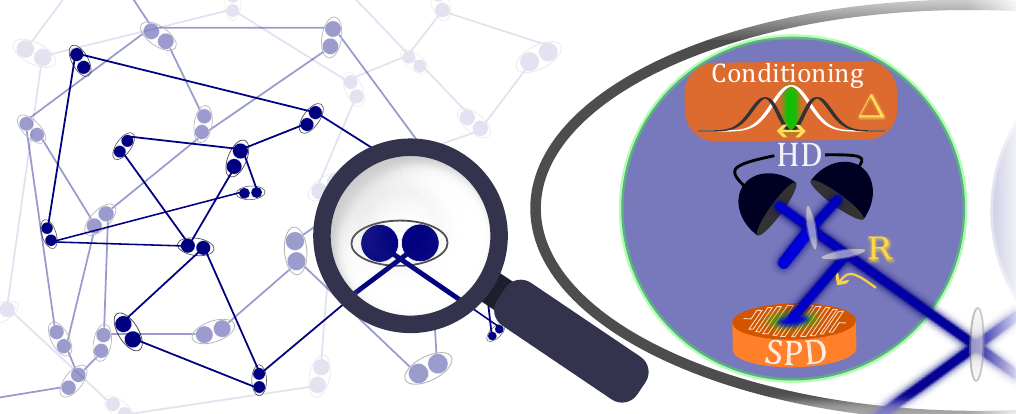}
\caption{Sketch of the hybrid Bell-state measurement (HBSM) consisting of an on-off single-photon detection (SPD) and a quadrature conditioning via homodyne detection (HD) in a network configuration. At a network node, two optical modes are mixed on a balanced beamsplitter. One of the output modes is tapped via a beamsplitter with reflectivity $R$ and this part is send to a SPD. The transmitted mode is directed to a HD, which is used for conditioning on quadrature values in a window $\Delta$ around $q=0$. The HBSM success is heralded by both detection events.}\label{fig1} 
\end{figure}

In this paper, we investigate a method to mitigate the detrimental effect of multi-photon components: the hybrid Bell-state measurement (HBSM). This approach combines on-off single-photon detection with homodyne conditioning, as illustrated in Fig.~\ref{fig1}. The addition of homodyne conditioning enables a more accurate discrimination between single- and two-photon states, reducing potential errors and thereby boosting the measurement projectivity. This hybrid approach has been used in our recent experimental realizations of heterogeneous swapping and of an optical qubit encoding converter based on entanglement between wave-like and particle-like qubits \cite{Morin2014,Guccione2020,Darras2023}, where it proved instrumental for successful demonstrations. Here, we provide the theoretical formalism and first thorough characterizations of the HBSM properties as a function of the tunable experimental parameters. We then apply this approach to the key examples of quantum teleportation and entanglement swapping protocols. We finally compare the performances with challenging photon-number resolving detection and show that the HBSM beats it in a very large range of parameters.


\section{Hybrid Bell-state measurement: formalism and error mitigation}

The hybrid Bell-state measurement is depicted in Fig.~\ref{fig1}. It combines an on-off single-photon detection (SPD) and quadrature conditioning via homodyne detection (HD). The incoming mode impinges on a beamsplitter with a reflectivity $R$ and the reflected path leads to the SPD. The transmitted mode is sent to a homodyne detection, where we define a conditioning window $\Delta$ around the quadrature value $q=0$. The success of the HBSM is heralded by both detection events. The HBSM has therefore two tunable parameters, the reflectivity $R$ and the conditioning window $\Delta$, and also depends on the detection efficiencies. In the following, $\Delta$ is normalized to the standard deviation of the vacuum shot noise. 

The general principle of the HBSM can be first understood in the limit of small $R$. In this case, the first detection is equivalent to a photon subtraction. A two-photon state then results in a single-photon state at the HD, while a single-photon state reduces to vacuum. A subsequent homodyne detection can discriminate the parity as the states have different probabilities of returning a given quadrature value. In particular, the probability of a value close to zero is large for vacuum and negligible for a single photon, as can be seen in Fig. \ref{fig1} where the associated white and black marginal distributions are shown. In the following, we provide the full theoretical analysis and benchmarking of the HBSM.  

\subsection{POVM derivation}

The overall detector can be described by its positive operator valued measure (POVM) \cite{Helstrom}. Each possible measurement outcome $x$ has a semipositive hermitian operator POVM element $\Pi^x$ which can be used to determine the probability of this outcome for a given impinging state $\rho$ as $\mathrm{Tr}[\Pi^x \rho]$. These operators sum up to the identity, $\sum_x \Pi^x=\mathbb{1}$. In the case of the HBSM, there are two possible measurement outcomes, corresponding to a successful and an unsuccessful event. Here we are interested in the successful POVM element $\Pi^{\mathrm{on}}$. 

The description of this POVM element relies on the matrices of the beamsplitter operation $\mathrm{BS}(R)$ that depends on its reflectivity, the single-photon detection POVM element $\Pi^{\mathrm{on}}_{\mathrm{SPD}}$, which corresponds here to an on-off detection \cite{DAuria}, and the homodyne detection POVM elements \cite{Leonhardt} (see Appendix \ref{App1}), integrated over the detection window $\Delta$ and independent of the detection phase $\theta$, i.e., 
\begin{eqnarray}
\Pi^{\mathrm{on}}_{\mathrm{HD}(\Delta)}&=&\int\limits_{-\Delta/2}^{\Delta/2} dq \int\limits_{0}^{2\pi} d\theta \; \Pi^{q,\theta}_{\mathrm{HD}}
\end{eqnarray} 
where $q$ is normalized to the vacuum noise. The $\Pi^{\mathrm{on}}$ HBSM matrix in the Fock basis can then be written as
\begin{eqnarray}
\Pi^{\mathrm{on}} = \mathrm{BS}(R)^{\dagger} [\Pi^{\mathrm{on}}_{\mathrm{HD}(\Delta)} \otimes \Pi^{\mathrm{on}}_{\mathrm{SPD}} ]\mathrm{BS}(R).
\end{eqnarray}

\begin{figure}[b!]
\includegraphics[width=0.95\columnwidth]{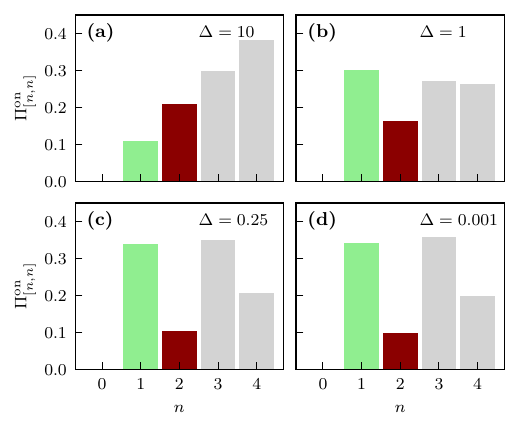}
\vspace{-0.2cm}
\caption{Diagonal elements of the HBSM POVM $\Pi^{\textrm{on}}$ element. They are given for a fixed reflectivity $R=0.1$ and for different conditioning windows $\Delta$, from a large acceptance to a very small one relative to shot noise: (a) $\Delta = 10$,  (b) $\Delta = 1$,  (c) $\Delta = 0.25$ and (d) $\Delta = 0.001$.  The coloring indicates the wanted green $\Pi^{\mathrm{on}}_{[1,1]}$ and unwanted red $\Pi^{\mathrm{on}}_{[2,2]}$ elements. All matrices were normalized for comparability, with normalization factors $\mathcal{N}_a=0.824$, $\mathcal{N}_b=0.156$, $\mathcal{N}_c=0.037$ and $\mathcal{N}_d=0.00015$. The detection efficiencies are set to $\eta_{\mathrm{HD}}~=~\eta_{\mathrm{SPD}}~=~0.9$. Diagonal elements $n>2$ are indicated in grey. In our study, their contributions will be negligible because the states to be measured have a very low probability of carrying more than two photons.}\label{fig2} 
\end{figure}

As this measurement is phase-independent, the POVM element has only diagonal terms that are non-zero. The full expressions of the first elements are given in Appendix \ref{App2}. As mentioned in the introduction we mostly assume components in the Fock basis up to $\ket{2}$. The diagonal elements of interest can be read as
\begin{eqnarray}
\label{POVM_math1}
    \Pi^{\mathrm{on}}_{[0,0]} &= &0 \nonumber \\
    \Pi^{\mathrm{on}}_{[1,1]} &= &R \; \eta_{\mathrm{SPD}} \;  \text{erf}\left(\frac{\Delta }{2}\right) \nonumber \\
     \Pi^{\mathrm{on}}_{[2,2]} &= &R \; \eta_{\mathrm{SPD}}  \Big[\text{erf}\left(\frac{\Delta }{2}\right) (2-R \;\eta_{\mathrm{SPD}})\nonumber \\
    & &+\frac{2  \Delta }{\sqrt{\pi }}\eta_{\mathrm{HD}} (R-1)e^{-\frac{\Delta ^2}{4}}\Big] 
\end{eqnarray}
where $\eta_{\mathrm{SPD}}$ is the detection efficiency of the single-photon detector and $\eta_{\mathrm{HD}}$ the efficiency of the homodyne detection. Many implementations require a filtering stage before the single-photon detection, and its efficiency can be integrated into an overall $\eta_{\mathrm{SPD}}$.

As the ideal HBSM shall project on the single-photon state, we aim at reducing $\Pi^{\mathrm{on}}_{[2,2]}$ while maximizing $\Pi^{\mathrm{on}}_{[1,1]}$ with the tunable parameters $R$ and $\Delta$. As shown in Fig. \ref{fig2}, for a fixed reflectivity $R$ this is obtained by decreasing the conditioning window $\Delta$. From a certain value of $\Delta$, typically 0.25, decreasing it further has a limited effect as can be seen in Fig. \ref{fig2}(c) and \ref{fig2}(d). Furthermore, we note that we are actually achieving a measurement that detects odd Fock states with higher probability than even ones, as expected from quadrature conditioning at the origin. In our case of interest for BSM, this feature reduces to a boosted single-photon versus two-photon detection, which can be more easily seen in the asymptotic limit of small $R$ and $\Delta$, where the POVM elements reduce to
\begin{eqnarray}
\label{POVM_math1lim}
    \Pi^{\mathrm{on}}_{[1,1]} &\sim &\frac{R \; \Delta  \; \eta_{\mathrm{SPD}} }{\sqrt{\pi }} \nonumber \\
     \Pi^{\mathrm{on}}_{[2,2]} &\sim &\frac{2\;R \; \Delta  \; \eta_{\mathrm{SPD}} }{\sqrt{\pi }}\left(1-\eta_{\mathrm{HD}}\right) .
\end{eqnarray}
To further explore the HBSM in this case a benchmark needs to be defined, as done in the following. 

\subsection{Projectivity benchmark: purity of the HBSM}
The ideal HBSM should act as a perfect single-photon detector, having only one non-zero element, i.e., $\Pi^{\textrm{on}}_{[1,1]}~=~1$. In order to benchmark the HBSM we propose the purity $\mathcal{P}( \Pi^{\textrm{on}})$, indicating how close the HBSM is to an ideal projective measurement. The POVM will be truncated such that we only consider the first diagonal elements defined in Eq. \ref{POVM_math1}. As above, this is due to the small higher order Fock components in the considered input states. The purity can then be defined as
\begin{eqnarray}
\mathcal{P}( \Pi^{\textrm{on}}) = \frac{ (\Pi^{\textrm{on}}_{[1,1]})^2+ (\Pi^{\textrm{on}}_{[2,2]})^2}{( \Pi^{\textrm{on}}_{[1,1]}+ \Pi^{\textrm{on}}_{[2,2]})^2}.
\end{eqnarray}  
This measure takes values between 0.5 and 1, where $0.5$ corresponds to $\Pi^{\textrm{on}}_{[2,2]}=\Pi^{\textrm{on}}_{[1,1]}$. In the asymptotic limit of $R\rightarrow 0$ and $\Delta \rightarrow 0$ the purity approaches
\begin{equation}
\mathcal{P}( \Pi^{\textrm{on}}) \sim \frac{1+ 4 (\eta_{\mathrm{HD}}-1)^2}{(3-2 \eta_{\mathrm{HD}})^2},\label{EqPurityHBSM}
\end{equation}
and only depends on the homodyne measurement efficiency $\eta_{\mathrm{HD}}$. For $\eta_{\mathrm{HD}} \rightarrow 1$,
the purity reaches unity, corresponding to a projective measurement with $\Pi^{\textrm{on}}_{[1,1]}=1$.

\begin{figure}[t!]
\includegraphics[width=0.95\columnwidth]{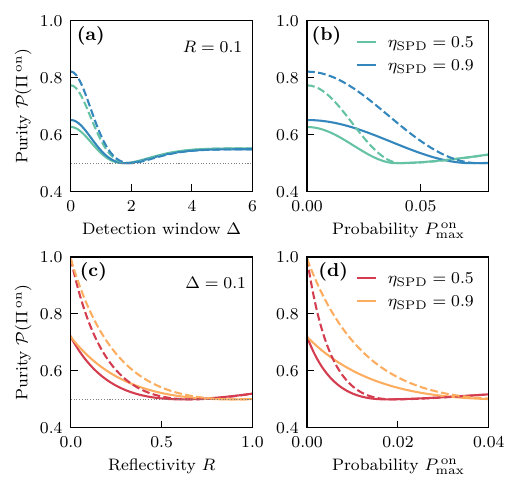}
\caption{HBSM purity $\mathcal{P}(\Pi^{\, \mathrm{on}})$. The purity is given as a function of (a) the conditioning window $\Delta$ with fixed reflectivity $R=0.1$ and of (c) the reflectivity $R$ with fixed window $\Delta = 0.1$. (b) and (d) provide the purity as a function of the detection probability $P_{\mathrm{max}}^{ \, \mathrm{on}}$. The homodyne detection efficiency is set to $\eta_{\textrm{HD}}=0.9$ ($\eta_{\textrm{HD}}=1$) for the solid (dotted) line, while two values of the single-photon detector efficiency $\eta_\textrm{SPD}$ are considered. The grey dotted line depicts the purity of a perfect on-off detector.}\label{fig3} 
\end{figure}

Given this expression, we can now evaluate how the purity depends on $\Delta $ and $R$. This is detailed in Fig. \ref{fig3}(a) and \ref{fig3}(c), respectively, and for two values of SPD detection efficiency $\eta_{\mathrm{SPD}}$. Figures \ref{fig3}(b) and \ref{fig3}(d) provide the maximal possible detection probability $P_{\mathrm{max}}^{ \, \mathrm{on}}$, which is defined as the maximal detection probability over all states \textbf{$P_{\mathrm{max}}^{ \, \mathrm{on}} = \max_{\rho}\mathrm{Tr}[\Pi^{\mathrm{on}}\rho]$}, thereby providing the detection probability for the most favorable state. The reflectivity $R$ has a larger impact on the purity while $\Delta$ shows a more favorable trade-off between purity and detection probability. We note also that a decrease in the SPD efficiency increases the importance of the two-photon component.

The purity for perfect homodyne detection efficiency and small $R$ and $\Delta$ can reach unity. Although for realistic values the purity of the HBSM decreases, it is still above the value of ideal on-off detectors. In the case of Bell-sate measurement, even small gains in purity can strongly boost the measurement. In fact, with $R$ and $\Delta$ well set, this detector can herald with high probability a single photon and suppress the detection of two photons. In the next section, we turn to the output of this hybrid measurement approach when applied to different input states and detail its use for networking protocols.  

\section{Application to the hybrid Bell-state Measurement}

\begin{figure}[t!]
\includegraphics[width=0.96\columnwidth]{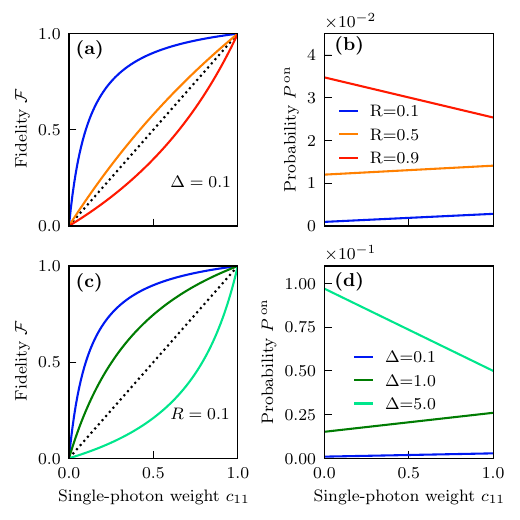}
\caption{Effect of the HBSM applied to one mode of a two-mode state with photon-number correlations, of the form $\sqrt{c_{11}}\ket{11}+\sqrt{1-c_{11}}\ket{22}$ or $c_{11}\ket{11}\bra{11}+(1-c_{11})\ket{22}\bra{22}$ where $c_{11}$ is the single-photon weight. The discrepancy from the ideal case, in which the HBSM projects the second mode onto $\ket{1}$, is given by the fidelity $\mathcal{F}$ as a function of the weight $c_{11}$.
The fidelity is displayed for (a) different reflectivities $R$ with fixed conditioning window $\Delta = 0.1$ and for (c) different conditioning windows $\Delta$ with fixed $R = 0.1$. The black dotted lines indicate the fidelity without any measurement. (b) and (d) provide the corresponding success probabilities $P^{\mathrm{on}}$. For the blue case in (d), the probability for $c_{11}=0.5$ is equal to $2.10^{-3}$. The detection efficiencies are set to $\eta_{\mathrm{HD}}=0.9$ and $\eta_{\mathrm{SPD}} = 0.5$. }\label{fig4}
\end{figure}

In the following, we will consider the general scenario of a two-mode input state and the HBSM applied on one of the two modes. The performance of the HBSM can be characterized by calculating the fidelity $\mathcal{F}$ of the conditioned mode with the ideal projection if a perfect single-photon measurement would be performed. We will first consider the case of an input state with single- and two-photon components, and then consider the use of the HBSM for two typical examples, namely quantum teleportation and swapping.

\subsection{The HBSM as a single-photon detector} 
We will first investigate the effect of the HBSM on two-mode states with photon-number correlations, i.e. of the form $\sqrt{c_{11}}\ket{11}+\sqrt{1-c_{11}}\ket{22}$ or  $c_{11}\ket{11}\bra{11}+(1-c_{11})\ket{22}\bra{22}$. These two cases are identical as the HBSM is phase-insensitive. An ideal measurement on one mode will project the other one onto a single-photon state $\ket{1}$. Specifically, we want to evaluate how the fidelity $\mathcal{F}$ between the conditioned mode and the target single photon depends on the HBSM settings, i.e., $\Delta$ and $R$. 

\begin{figure}[t!]
\includegraphics[width=0.94\columnwidth]{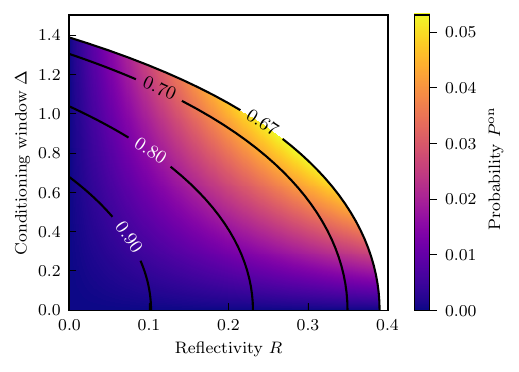}
 \caption{Detailed map of the HBSM success probability $P^{\mathrm{on}}$ for an impinging two-mode state with a single-photon weight $c_{11}=0.5$. $P^{\mathrm{on}}$ is given as a function of the conditioning window $\Delta$ and the reflectivity $R$. Isolines correspond to fidelity values. The detection efficiencies are set to $\eta_{\mathrm{HD}}=0.9$ and $\eta_{\mathrm{SPD}} = 0.5$.} \label{fig5}
\end{figure}

Figure \ref{fig4} provides the fidelity $\mathcal{F}$ and the associated success probability $P^{\mathrm{on}}$ of the measurement as a function of the single-photon component $c_{11}$. Figures \ref{fig4}(a) and \ref{fig4}(b) correspond to a given conditioning window and different reflectivities. A large reflectivity $R\geqslant 0.5$ results in a projection worse than simply tracing out the second mode without a measurement. As expected, the success probability $P^{\mathrm{on}}$ is increasing with the single-photon weight, except when the reflectivity is very large (here $R=0.9$), due to SPD events corresponding more and more to a two-photon detection.

\begin{figure*}[t!]
\includegraphics[width=1.8\columnwidth]{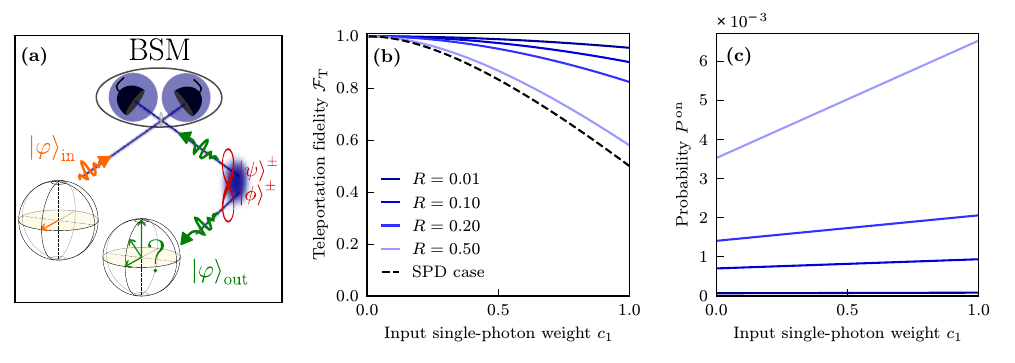}
\caption{Teleportation protocol and hybrid Bell-state measurement. (a) A qubit to be teleported $\ket{\varphi}_{in}=\sqrt{c_0}\ket{0}+\sqrt{c_1}\ket{1}$ undergoes a BSM with one mode of a Bell state. The remaining mode is ideally projected onto the input state upon successful detection. (b) The fidelity $\mathcal{F}_{\mathrm{T}}$ between the input and teleported states and (c) the success probability $P^{\mathrm{on}}$ are given as a function of the input single-photon weight $c_1$ for different reflectivities $R$ of the HBSM.  The black dashed line indicates the SPD case, i.e., using an on-off detector with perfect efficiency, and without homodyning. The conditioning window is fixed at $\Delta = 0.1$ and the detection efficiencies are set to $\eta_{\mathrm{HD}}=0.9$ and $\eta_{\mathrm{SPD}} = 0.5$. }\label{fig6Tele} 
\end{figure*}

In Fig. \ref{fig4}(c) and \ref{fig4}(d) the reflectivity is fixed to $R~=~0.1$ while the conditioning window $\Delta$ varies. It can be observed that a large conditioning window has a stronger detrimental effect than a high reflectivity seen. This leads to the conclusion that the conditioning window is the more critical parameter, whereas the reflectivity can act as fine-tuning. This effect strongly depends on the SPD efficiency $\eta_{\mathrm{SPD}}$, which here is set to $\eta_{\mathrm{SPD}}=0.5$. 

The tuning of both $R$ and $\Delta$ enables to achieve a large fidelity with the target single photon but reduces the success probability $P^{\mathrm{on}}$. This trade-off is shown in Fig. \ref{fig5}, where the success probability is given as a function of the reflectivity and of the conditioning window. At the cost of a low success probability, arbitrarily high fidelities can be achieved, even if the SPD efficiency is set as here to a moderate value $\eta_{\mathrm{SPD}}=0.5$.

As we have shown, for some sets of parameters, the HBSM is thereby able to efficiently discriminate a two-photon contribution from a single-photon one, at the cost of the success probability. For linear BSM the usual detection schemes do not enable this discrimination and two-photon errors are one of the limiting factors, hindering faithful quantum teleportation and entanglement swapping. We will now focus on the performances when a HBSM is used for these two operations.

\subsection{The HBSM in a teleportation protocol}
Quantum teleportation as shown in Fig. \ref{fig6Tele}(a) considers an input qubit of the form $\ket{\varphi}_{\mathrm{in}}~=~\sqrt{c_0}~\ket{0}~+~\sqrt{c_1}\ket{1}$ and one of the four entangled Bell states $\ket{\phi}^\pm~=~(1/ \sqrt{2})(\ket{00} \pm \ket{11})$ or $\ket{\psi}^\pm = (1/ \sqrt{2})(\ket{01} \pm \ket{10})$ in the photon-number basis as a resource. A Bell-state measurement between one mode of the Bell state and the input state will then project the remaining mode into the input state up to a local unitary, as reminded here \cite{Chuang}:

 \begin{alignat}{5}
\ket{\varphi}_{\mathrm{in}}\otimes \ket{\psi}^+ \propto \; &\ket{\psi}^+ \otimes &\ket{\varphi}_{\mathrm{out}} &+ &\ket{\psi}^- \otimes &Z \ket{\varphi}_{\mathrm{out}} +  \nonumber \\ 
 &\ket{\phi}^+ \otimes &X \ket{\varphi}_{\mathrm{out}} &+  &\ket{\phi}^- \otimes &ZX \ket{\varphi}_{\mathrm{out}}.\label{TeleMathBeforeBS}
 \end{alignat}
 
 Any linear BSM consists of a balanced beamsplitter after which one detector can be placed at each output arm:

 \begin{eqnarray}
\mathrm{BS}(0.5)\ket{\psi}^+ &=& \; \ket{01} \nonumber \\
\mathrm{BS}(0.5)\ket{\psi}^- &= &\; \ket{10} \nonumber \\ 
\mathrm{BS}(0.5)\ket{\phi}^{\pm} &=& \; \frac{1}{\sqrt{2}}\ket{00}\pm \frac{1}{2}(\ket{02}-\ket{20}),  \label{TeleMathAfterBS}
\end{eqnarray}
Only the states $\ket{\psi}^{\pm}$ can be detected with this scheme and this detection relies on the fact that one can detect exactly a single photon. For simplification, we will only consider $\ket{\psi}^+$. The probability to measure a single photon is $0.25$, while the probability to measure two photons is $c_1 / 4$, and therefore depends on the single-photon weight $c_1$ of the input qubit. In the usual BSM scheme, an on-off detector is used. As it detects single-photon and two-photon components equally, we can define the worst case scenario of the teleportation fidelity $\mathcal{F}_{\mathrm{T}}$ by using an on-off detector with perfect efficiency.

Figures \ref{fig6Tele}(b) and \ref{fig6Tele}(c) provide the teleportation fidelity $\mathcal{F}_{\mathrm{T}}$ and the probability of success when using a HBSM, for a fixed conditioning window and various values of the reflectivity $R$. The HBSM leads to fidelities larger than the ones obtained with the usual BSM for all possible single-photon input weights. 

 \begin{figure*}[t!]
\includegraphics[width=1.8\columnwidth]{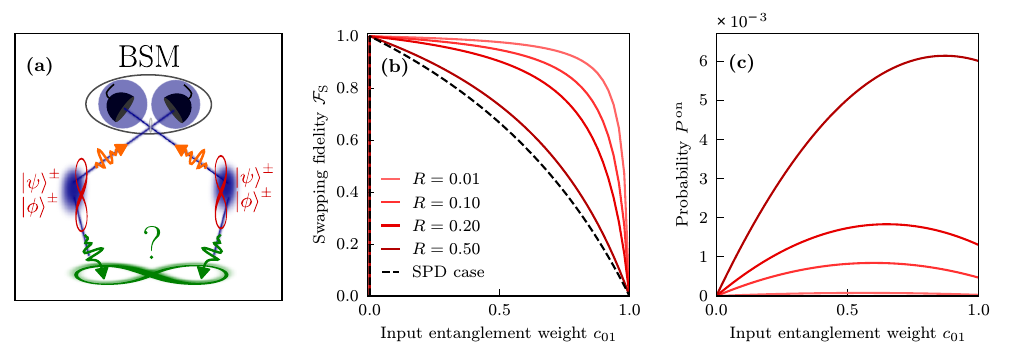}
\caption{Swapping protocol and hybrid Bell-state measurement. (a) Two Bell states of the form $\ket{\psi^{\textrm{in}}} =  \sqrt{c_{01}}\ket{01}+\sqrt{c_{10}}\ket{10}$ are created and the second mode of each undergoes a BSM. This operation ideally entangles the two remaining modes onto $\ket{\psi^{\textrm{out}}}\propto \ket{01}+\ket{10}$. (b) The fidelity $\mathcal{F}_{\mathrm{S}}$ between the ideal $\ket{\psi^{\textrm{out}}}$ and the projected state and (c) the success probability $P^{\mathrm{on}}$ are given as a function of the input weight $c_{10}$ for different reflectivities $R$ of the HBSM. The coincidence window is fixed at $\Delta = 0.1$ and the detection efficiencies are set to $\eta_{\mathrm{HD}}=0.9$ and $\eta_{\mathrm{SPD}} = 0.5$.}\label{fig7Swap}
\end{figure*}

\subsection{The HBSM in a swapping protocol}

We now turn to the case of entanglement swapping, as sketched in Fig. \ref{fig7Swap}(a). Here the input states are twice the same Bell state. A Bell state measurement between one mode of each then projects the remaining modes once again into a Bell state. We assume that the two input entangled states are of the form $\ket{\psi}_{\textrm{in}}^{\sqrt{c_{10}}}~=~\sqrt{c_{01}}\ket{01}~+~\sqrt{c_{10}}\ket{10}$ such that their tensor product can be written as: 
\begin{align}
\ket{\psi}_{\textrm{in}}^{\sqrt{c_{10}}}\otimes \ket{\psi}_{\textrm{in}}^{\sqrt{c_{10}}} \propto \; &\ket{\psi}^+ \otimes \sqrt{c_{01}c_{10}}(\ket{01}+\ket{10})+ \nonumber \\ &\ket{\psi}^- \otimes \sqrt{c_{01}c_{10}}(\ket{10}-\ket{01}) +  \nonumber \\ 
 &\ket{\phi}^+ \otimes (c_{01}\ket{00}+c_{10}\ket{11}) + \nonumber \\ &\ket{\phi}^- \otimes (c_{10}\ket{11}-c_{01}\ket{00}),\label{SwapMathBeforeBS} 
\end{align}
where the BSM is applied between each second mode. After one mode of each interacted on a balanced beamsplitter, a projection onto a single-photon state will lead to a Bell state $\ket{\psi}^\pm$, dependent on the detector. We will again consider the detection on only one output port, projecting the final modes into $(1/\sqrt{2})(\ket{01}+\ket{10})$. Equal detection of single photons with probability $c_{01}(1-c_{01})$ and two photons $ 0.5 c_{01}^2$ leads to the worst case of the swapping fidelity $\mathcal{F}_{\mathrm{S}}$ between the ideal output state $\ket{\psi^{\textrm{out}}}$ and the actual output state, defined by the use of an ideal on-off detector.

Figure \ref{fig7Swap}(b) and \ref{fig7Swap}(c) provide the swapping fidelity $\mathcal{F}_{\mathrm{S}}$ and the probability of success when using a HBSM, for a fixed conditioning window and various values of the reflectivity $R$. Similarly to the teleportation case, the HBSM leads to fidelities larger than the ones obtained with the usual BSM for all possible single-photon input weights.

We note that the swapping fidelity drops to zero for the extreme cases $c_{01}=0$ and $c_{01}=1$. Compared to the teleportation fidelity of Fig. \ref{fig6Tele}(b) this might seem peculiar at first glance. However, here entanglement between two remote states is verified, which cannot be created at all if the input states are not entangled, i.e., at the extreme points of $c_{01}$. For decreasing $c_{01}$, the swapping state becomes closer to the target state $\ket{\psi}^+$ as the two photon detection probability is decreasing steadily. This improvement in fidelity drops abruptly to zero for $c_{01}=0$, corresponding to no entanglement in the system that could be swapped. This differs from teleportation where the theoretical minimum is not zero. Depending on the reflectivity $R$, the two-photon event probability reduces slower or faster. Those false two-photon detection events shift the maximum of the success probability $P^{\mathrm{on}}$ towards higher input weights $c_{01}$ as shown in Fig. \ref{fig7Swap}(c).

We have seen that the HBSM enables substantially improved output state fidelity of both swapping and teleportation operations. We will now compare the HBSM to photon-number resolving (PNR) detectors, which are another option to improve linear BSM. 

\section{Benchmarking the HBSM against photon-number resolving detectors}

\begin{figure}[b!]
\includegraphics[width=0.85\columnwidth]{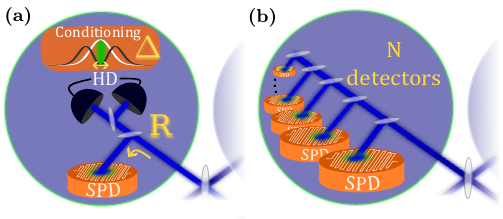}
\caption{Sketch of Bell-state measurement implementations: (a) with a single-photon on-off detector in a HBSM scenario as studied before and (b) with $N$ multiplexed single-photon on-off detectors with equally distributed splitting. This multiplexed implementation leads to photon-number resolution in the limit of a large number $N$ of detectors. We compare the two configurations.}\label{fig8} 
\end{figure} 

An obvious but daunting solution to two-photon errors are photon-number resolving detectors. Those detectors are under intense development, and promising candidates are transition-edge sensors or superconducting-nanowire detectors that are multiplexed or where the number resolution is based on different dynamics \cite{Lita2022}. There is a maximum number of $N$ photons that can be distinguished in realistic detectors \cite{Lita2008,Lita2022}. 

Independent of the technology, all PNR detectors can be described via the same underlying POVM elements \cite{Provaznik2020, Cohen2018}, equivalent to $N$ multiplexed detectors. In order to compare PNR detectors to the HBSM, as sketched in Fig. \ref{fig8}, we will only consider the POVM element of a single detection event with a photon resolution up to $N$, $\Pi^{\mathrm{(1,N)}}$. All PNR POVM elements are phase-independent such that their matrices only have diagonal elements and can be written as:
\begin{eqnarray}
 \Pi^{\mathrm{(1,N)}}_{[0,0]} &= &0 \\
 \Pi^{\mathrm{(1,N) }}_{[1,1]} &= &\eta_{\mathrm{SPD}} \nonumber\\
 \Pi^{\mathrm{(1,N) }}_{[2,2]} &= &\eta_{\mathrm{SPD}} \left[2\left(1-\eta_{\mathrm{SPD}}\right)+\frac{\eta_{\mathrm{SPD}}}{N}\right].\nonumber
 \label{PNR_POVM2}
\end{eqnarray}

\begin{figure}[t!]
\includegraphics[width=0.97\columnwidth]{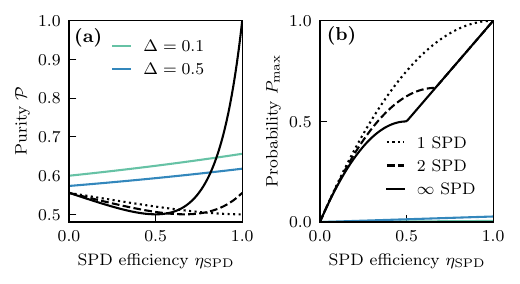}
\caption{Comparison of the hybrid BSM with BSM implementations based on N-multiplexed single-photon detectors. The POVM success element of the HBSM $\Pi^{\mathrm{on}}$ is benchmarked against the POVM elements of N-multiplexed single photon detectors $\Pi^{\mathrm{(1,N)}}$. The purity $\mathcal{P}$ is given in (a) as a function of the SPD efficiency $\eta_{\mathrm{SPD}}$. The corresponding success probability $P_{\mathrm{max}}$ is given in (b). Two values of conditioning window $\Delta$ are considered for HBSM, and compared to implementations with $N=1$, $N=2$, and $N=\infty$ that emulates perfect photon-number resolution. All POVM elements are truncated at $n=2$. For HBSM, the homodyne detection efficiency is set to $\eta_{\mathrm{HD}}=0.9$, and the reflectivity is fixed at $R=0.1$.} \label{fig9} 
\end{figure}

We note that even in the limit of perfect PNR $N \rightarrow \infty$ imperfect detection efficiency $\eta_{\mathrm{SPD}} <1$ will lead to an unavoidable two-photon error with
\begin{eqnarray}
 \lim_{N \rightarrow \infty}\Pi^{\mathrm{(1,N) }}_{[2,2]}=2\eta_{\mathrm{SPD}}(1-\eta_{\mathrm{SPD}}).
 \end{eqnarray}
This trend is to be compared to the HBSM in Fig. \ref{fig9}(a), where the purity of either the HBSM POVM element $ \Pi^{\mathrm{on}}$ or the PNR POVM element $ \Pi^{\mathrm{(1,N)}}$ are shown as a function of $\eta_{\mathrm{SPD}}$. We can observe that for a PNR detector with $N=1$ the HBSM with small $\Delta$ has a higher purity for all $\eta_{\mathrm{SPD}}$. This is indeed also true for $N=2$ even if here we can observe that towards larger SPD efficiency the purity recovers, still staying below the one of the HBSM.

Remarkably, even the ideal PNR ($N \rightarrow \infty$) only surpasses the HBSM purity with $\Delta = 0.1$ from $\eta_{\mathrm{SPD}}=0.85$ onward, a very stringent condition. Indeed if we compare the limits of the HBSM purity of eq. \eqref{EqPurityHBSM}, we can see that it is independent of the efficiency $\eta_{\mathrm{SPD}}$, while the purity of the PNR depends on it as
\begin{eqnarray}
 \lim_{N \rightarrow \infty}\mathcal{P}(\Pi^{\mathrm{(1,N)}}_{[1,1]}) = \frac{4 \eta_{\mathrm{SPD}}^2-8 \eta_{\mathrm{SPD}}+5}{(3-2 \eta_{\mathrm{SPD}})^2}<1  \quad\textrm{for all $\eta_{\mathrm{SPD}}$}.\nonumber
 \end{eqnarray}
This explains the strong advantage of the HBSM in terms of purity. As stressed before, it comes at the cost of a low success probability, as shown in Fig. \ref{fig9}(b).

\begin{figure}[t!]
\includegraphics[width=0.97\columnwidth]{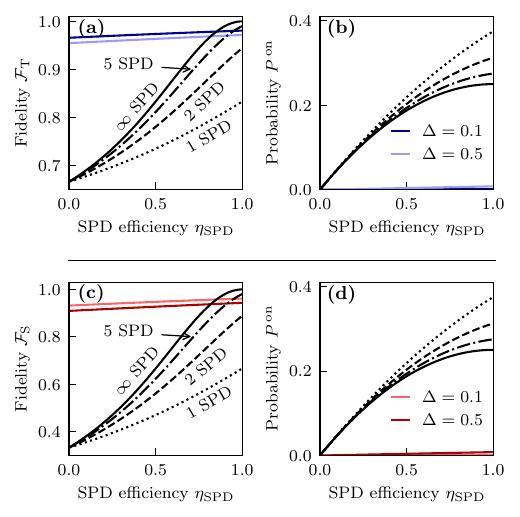}
\caption{Comparison of BSM implementations for quantum teleportation and entanglement swapping. (a) and (b) provide the fidelity $\mathcal{F}_{\mathrm{T}}$ and success probability $P^{ \, \mathrm{on}}$ for the teleportation of a balanced qubit. (c) and (d) provide the fidelity $\mathcal{F}_{\mathrm{S}}$ and success probability $P^{ \, \mathrm{on}}$ for the swapping of two balanced $\ket{\phi^+}$ states. Two values of conditioning window $\Delta$ are considered for HBSM, and compared to implementations with $N=1$, $N=2$, $N=5$, and $N=\infty$ that emulates perfect photon-number resolution. All POVM elements are truncated at $n=2$. For HBSM, the homodyne detection efficiency is set to $\eta_{\mathrm{HD}}=0.9$, and the reflectivity is fixed at $R=0.1$.}\label{fig10}  
\end{figure}

We further want to compare the performance of the HBSM and PNR detectors in the typical use cases of teleportation and entanglement swapping. The results are summarized in Fig. \ref{fig10}. Here all states are balanced, including the teleportation input state $(1/\sqrt{2})(\ket{0}+\ket{1})$ and the swapping input states $(1/\sqrt{2})(\ket{00}+\ket{11})$. The achieved fidelities with the HBSM with $\Delta = 0.1$ are superior to the ones obtained with the ideal PNR detector with $N=\infty$ until an efficiency $\eta_{\mathrm{SPD}}=0.79$ in the case of teleportation and $\eta_{\mathrm{SPD}}=0.86$ in the case of swapping. Up to typically $N=5$, the fidelities are always larger for the HBSM whatever the efficiency. These numbers illustrate the advantage that the HBSM can provide in a variety of scenarios. 
 
Indeed PNR detectors are not largely commercially available at the moment, and  the threshold value we obtain is comparable to the best values demonstrated for PNR detectors, reachable at a limited number of wavelengths.

\section{Conclusion}
In summary, we have presented a novel hybrid Bell-state measurement, combining homodyne conditioning with single-photon detection. This original BSM implementation effectively mitigates the two-photon errors in this measurement, thereby significantly enhancing quantum teleportation and entanglement swapping fidelities. This advantage comes at the cost of a reduced success rate due to the conditioning process. We have provided a comprehensive study of the HBSM properties as a function of the tunable experimental parameters, including detection efficiencies, tapping ratio and conditioning window. The effect of these parameters on protocol fidelities has been characterized. Notably, we have also found that this strategy outperforms in fidelity performances ideal photon-number resolving detectors for efficiencies up to typically 90\%, a very challenging value. This study confirms the relevance of combining detection methods and calls for a broader exploration of the capabilities opened by hybrid detection schemes in the realm of linear-optical quantum information processing. 

\begin{acknowledgments}
The authors acknowledge early discussions with A.~Cavaill\`es, G. Guccione and H. Le Jeannic. This work was supported by the French National Research Agency via the ShoQC Quantera project (ANR 19-QUAN-0005-05) and France 2030 project QCommTestbed (ANR-22-PETQ-0011), and by the European Union's Horizon Europe research and innovation programme via the QIA-Phase 1 project (101102140). This work was also partly supported by Japan Science and Technology (JST) Agency (Moonshot R \& D) Grant No. JPMJMS2064, UTokyo Foundation, and donations from Nichia Corporation of Japan. A.K. acknowledges financial support from the Forefront Physics and Mathematics Program to Drive Transformation (FoPM) and Leadership Development Program for Ph.D. (LDPP), the University of Tokyo. A.B. acknowledges the support from the France 2030 project QuanTEdu-France. J.L. is a member of the Institut Universitaire de France.
\end{acknowledgments}

\appendix
\section{Definition of the homodyne POVM}\label{App1}
\vskip-1em
As used in our HBSM POVM derivation, here we remind the POVM element corresponding to a homodyne detection \cite{Leonhardt} 
\begin{eqnarray}
    \text{HD}(x,\theta) &=& \ket{x,\theta}\bra{x,\theta}  \nonumber\\
   \ket{x,\theta}&=& \expec{x_{\theta}|n}\nonumber \\
   &=&e^{in\theta}(\sqrt{\pi \nq} \; 2^n \;n!)^{(-1/2)}e^{-x^2/(2\nq)}H_n(\frac{x}{\sqrt{\nq}})\nonumber,
\end{eqnarray}
which is the wavefunction of Fock states projected onto the quadrature $x_{\theta}$. We can account for an imperfect detection efficiency $\eta$, by applying a beamsplitter operation, called Bernoulli transformation $B$. The resulting homodyne POVM elements can be written as 
\begin{eqnarray}
    \Pi_{\eta}(x,\theta) &=& \sum_{m,n,i} B_{m+i,m}(\eta) \; B_{n+i,n}(\eta)\;\nonumber\\ 
    &&\qquad\qquad\text{HD}(x,\theta)\ket{n+i}\bra{m+i}
    \label{HD_measurementOperator} \nonumber\\
    B_{m+i,m} &=& \sqrt{\binom{m+i}{m}\eta^m(1-\eta)^{i}}\nonumber.
\end{eqnarray}
We can therefore interpret this measurement as sampling from the marginal distribution along $x_{\phi}$ of the Wigner function.
\section{HBSM POVM elements}\label{App2}
We provide the expression for the HBSM POVM elements up to $n=4$ as displayed in Fig.~\ref{fig2}.
\begin{eqnarray}
    \Pi^{\mathrm{on}}_{[0,0]} &= &0 \nonumber \\
    \Pi^{\mathrm{on}}_{[1,1]} &= &R \; \eta_{\mathrm{SPD}} \;  \text{erf}\left(\frac{\Delta }{2}\right) \nonumber \\
     \Pi^{\mathrm{on}}_{[2,2]} &= &R \; \eta_{\mathrm{SPD}}  \Big[\text{erf}\left(\frac{\Delta }{2}\right) (2-R \;\eta_{\mathrm{SPD}}) \nonumber \\
    & &+\frac{2  \Delta }{\sqrt{\pi }}\eta_{\mathrm{HD}} (R-1)e^{-\frac{\Delta ^2}{4}}\Big] \nonumber \\
    \Pi^{\mathrm{on}}_{[3,3]} &= &R\; \eta_{\mathrm{SPD}} \; \text{erf}\left(\frac{\Delta }{2}\right) (\eta_{\mathrm{SPD}}\; R (\eta_{\mathrm{SPD}}\; R-3)+3) \nonumber \\
    & &-\frac{1}{4 \sqrt{\pi }}3\; e^{-\frac{\Delta ^2}{4}} \Delta \; \eta_{\mathrm{SPD}} \eta_{\mathrm{HD}} (R-1) R  \nonumber \\  
    & &\times \Big[\left(\Delta ^2-6\right) \eta_{\mathrm{HD}} (R-1)+4\; \eta_{\mathrm{SPD}}\; R-8\Big] \nonumber \\  
    \Pi^{\mathrm{on}}_{[4,4]} &= &\frac{1}{6} \eta_{\mathrm{SPD}}\; R \Biggl(\frac{1}{\sqrt{\pi }}
    e^{-\frac{\Delta ^2}{4}} \Delta \;  \eta_{\mathrm{HD}} (R-1) \Big[9 \left(\Delta ^2-6\right)  \nonumber \\
   & &\times \eta_{\mathrm{HD}} (R-1) (\eta_{\mathrm{SPD}} \; R-2)+\left(\Delta ^4-20 \Delta ^2+60\right)\nonumber \\
   & &\times \eta_{\mathrm{HD}}^2 (R-1)^2+24 \; \eta_{\mathrm{SPD}} \; R (\eta_{\mathrm{SPD} \;} R-3)+72\Big] \nonumber \\
   & &-6 \; \text{erf}\left(\frac{\Delta }{2}\right) (\eta_{\mathrm{SPD}} \; R-2) \nonumber  \\ 
   & &\times (\eta_{\mathrm{SPD}} \; R (\eta_{\mathrm{SPD}} \; R-2)+2)\Biggr)\nonumber.
\end{eqnarray}
\pagebreak

\end{document}